\documentclass[11pt, a4paper]{article}
\usepackage{amsmath,amssymb,amsfonts}
\usepackage[utf8]{inputenc}
\newcommand{\Adv}{\mathop{\mathrm{Adv}}}
\newcommand{\negl}{\mathop{\mathrm{negl}}}

\newcommand{\AdvPRG}[1]{\mathsf{Adv}_{\mathrm{PRG}}{#1}}

\newcommand{\AdvVCDH}[1]{\mathsf{Adv}_{\mathrm{V-CDH}}{#1}}
\usepackage[T1]{fontenc}
\usepackage{lmodern}
\usepackage{graphicx}
\usepackage{amsthm}
\usepackage[numbers]{natbib}
\usepackage{array}     
\usepackage{multirow}  
\usepackage{caption} 
\usepackage{booktabs}
\usepackage{algorithm}
\usepackage{algpseudocode} 
\usepackage{hyperref}
\usepackage{geometry}
\begin{document}
   \title{Primitive Vector Cipher(PVC):
   A Hybrid Encryption Scheme based on the Vector Computational Diffie–Hellman (V-CDH) Problem}
   
\author{\textbf{Gülçin Çivi Bilir}\\Istanbul Technical University \\Department of Mathematics Engineering  \\ civi@itu.edu.tr}
\date{}
\maketitle 

\begin{abstract}

This work introduces the Primitive Vector Cipher (PVC), a novel hybrid encryption scheme integrating matrix-based cryptography with advanced Diffie–Hellman key exchange. PVC's security is grounded on the established hardness of the Vector Computational Diffie–\-Hellman (V-CDH) problem. The two-layered design uses HKDF to mask plaintext via a DH-authenticated shared primitive vector and randomize cipher blocks with a per-block offset. This approach eliminates deterministic repetitions and provides strong resistance against linear and known-plaintext attacks. PVC’s block-wise structure allows for massive parallelism and excellent linear scaling. Security is formally analyzed, demonstrating IND-CPA security under V-CDH. STS protocol integration elevates security toward IND-CCA guarantees.
\end{abstract}

			\section{Introduction}\label{sec1}
	
	 With the rapid advancement of computer technology, digital platforms have become integral not only to our daily lives but also to political, economic, sociocultural, healthcare, and legal systems. 
	This multifaceted reliance on digital systems necessitates the protection, storage, transmission, and processing of personal, corporate, and institutional data from unauthorized third parties, thereby driving ongoing research in cryptographic systems. 
	To address these security concerns, many researchers have developed reliable, practical cryptographic systems to ensure the confidentiality, integrity, and authentication of data in a fast and secure manner.
	
	Cryptosystems are fundamentally classified into two main categories: symmetric and asymmetric systems. 
	In symmetric systems, the same key (referred to as the secret key) is used for both encryption and decryption of data. In contrast, asymmetric systems employ distinct public and private keys; while they resolve the key distribution problem inherent in symmetric systems, they are generally more computationally intensive.

In 1976, Diffie and Hellman introduced the concept of asymmetric key cryptography in their seminal paper, \emph{New Directions in Cryptography}, presenting what is now known as the Diffie–Hellman protocol—a secure method for key exchange \cite{diffie1976}. Its security rests on the computational hardness of the Discrete Logarithm Problem (DLP), enabling two parties to establish a fresh shared secret for each communication session and ensuring forward secrecy. Theoretical foundations and modern cryptanalytic approaches have been comprehensively examined by Maurer and Wolf \cite{Maur2000}. Although widely adopted due to its simplicity and robustness, the basic protocol is susceptible to active interception when authentication is absent. While the Diffie–Hellman protocol provides a cornerstone of asymmetric cryptography, algebraic structures such as matrices play a central role in both symmetric and asymmetric cryptosystems. Matrix-based encryption methods were first introduced into cryptography by Hill in 1929. Since then, extensions of the Hill cipher and other matrix-based approaches have been incorporated into modern protocols and inspired many novel designs.

Building on this foundation, we propose the \text{\emph{Primitive Vector Cipher}} (PVC), an asymmetric encryption scheme that integrates DH–STS–based key exchange with structured submatrix encryption, designed to deliver high entropy, strong statistical security, and resistance to linear, differential, and algebraic attacks.

The main contributions of this work are summarized as follows:

	\begin{enumerate}
		\item We introduce the notion of a \emph{primitive vector} $\mathbf{g}=(g_1,g_2,g_3) \in (\mathbb{F^*}_p)^3\bmod p$, in which each component is a distinct primitive root modulo a large prime $p$. This structure forms the basis for secret key generation and is the foundation for the \textbf{Vector Computational Diffie-Hellman ($\text{V-CDH}$)} problem.

 \item Encryption is carried out in two stages: 
 \vspace{.2cm}

        \textbf{Stage 1 – Submatrix encryption.} Each selected \(3 \times 3\) submatrix is first 
extracted from the masked master matrix via HKDF-seeded CSPRNG. These masked submatrices are 
then encrypted with the secret key matrices $\mathbf{V}$ and $\mathbf{U}$.
		\vspace{.2cm}
        
 \textbf{Stage 2 – Vector offset.} The column vectors of each encrypted block are further masked by a keystream derived via HKDF (HMAC–SHA-256) in counter mode \cite{RFC5869},\cite{NISTMD}, \cite{Shukla}. This two-layer design prevents deterministic repetitions, maintains ciphertext freshness, and, crucially, provides a strong {theoretical defense against linear and Known-Plaintext Attacks (KPA)} by obfuscating the core linear transformation.

 \item Unlike the Hill cipher and existing matrix-based schemes, our approach combines sub\-mat\-rix encryption with two secret matrices, \( \mathbf{V} \) and \( \mathbf{U} \) generated through the Diffie–Hellman protocol and the primitive vector \( \mathbf{g} \in (\mathbb{F^*}_p)^3 \).
		
\item The plaintext is first mapped to an ASCII matrix and masked with a random matrix $\mathbf{R}$ generated by a CSPRNG seeded with a hash of the primitive vector. This significantly increases entropy and statistical unpredictability.

\item \textbf{Provable security.}  
		Security formally reduces to two standard assumptions: the pseudorandom-generator advantage ($\AdvPRG{q}$) and the difficulty of solving the {Vector Computational Diffie-Hellman (V-CDH)} problem.
        A hybrid argument shows that an adversary’s total advantage is at most $\AdvPRG{q}+\AdvVCDH{t}+\negl(\lambda)$. The combination of the $\text{V-CDH}$ foundation and the column-wise counter offset serves as an integrity check, thereby extending the reduction to {IND–CCA security}.
		
		\item Our scheme augments the Diffie–Hellman key exchange with the Station-to-Station (STS) protocol to provide explicit authentication and mitigate MITM attacks.
	\end{enumerate}

\section{Matrices in Cryptology and Related Works}\label{sec:related}

		Modern cryptography extends beyond merely ensuring secure communication; it is grounded in fundamental primitives that provide confidentiality, integrity, authentication, and data protection. Key agreement protocols such as Diffie–Hellman~\cite{diffie1976,Maur2000}, symmetric encryption algorithms, asymmetric encryption schemes, digital signatures, and cryptographically secure hash functions constitute the backbone of secure systems. Since Hill's introduction of the first matrix-based block cipher in 1929, matrices have remained integral to cryptographic design\cite{Hill1929}.
		
		Matrices have long played a fundamental role in both classical and modern cryptography, serving as the basis for block ciphers, key exchange protocols, and public–key cryptosystems~(\cite{mceliece1978},\cite{rsa1978}). In symmetric encryption, matrix–based ciphers offer high–throughput parallelization and well–structured diffusion, while in public–key cryptography, they provide a rich algebraic framework for constructing hard computational problems.  
		The Hill cipher, one of the earliest matrix–based schemes, encrypts plaintext blocks through multiplication by an invertible key matrix over a finite field. it established the foundation for modern matrix-based cryptography. Extensions of the Hill cipher, such as the scheme proposed by Thilaka and Rajalakshmi in 2005, aimed to improve its security by incorporating affine and polynomial transformations~\cite{thilaka2005}. Another notable variation is due to Saeednia, who introduced random permutations of matrix rows and columns to generate a dynamic key for each encryption, relying only on matrix products and efficient primitive operations~\cite{saeednia2000}. In 2012, Reddy et~al.\ presented a modified version of the Hill cipher based on circulant matrix structures, designed to enhance cryptographic strength~\cite{reddy2012}. Despite these advances, cryptanalysis by Gupta, Singh, and Chaudhary in 2007 demonstrated that many such variants remain vulnerable~\cite{gupta2007}.

		  Nevertheless, matrix-based cryptographic constructions not only underpinned early schemes such as the Hill cipher, but also continue to play a crucial role in contemporary primitives, most notably the AES block cipher. In this sense, matrix techniques establish a conceptual bridge between classical and modern cryptographic paradigms. Indeed, AES, standardized by Rijndael, dominates block cipher deployment due to its strong resistance to differential and linear cryptanalysis~\cite{rijndael1999,rijndaelbook}. Recent research extends matrix-based techniques into new domains, including Hill cipher constructions over Gaussian integers, public-key systems based on generalized Fibonacci matrices, and hybrid designs leveraging cryptographically secure pseudorandom number generators (CSPRNGs), chaotic masking, or steganographic techniques~\cite{saba2024,Panchal,stego,stego2}.

Beyond symmetric constructions, matrix-based methods have also been adapted to public-key settings. In particular, the Diffie--Hellman (DH) protocol, a cornerstone of public-key cryptography, has inspired a wide range of matrix-based generalizations, including batch key agreement schemes, non-commutative ring extensions, and constructions involving high-order matrix exponentiation and primitive Galois matrices \cite{yacobi1997,etfektari2012,beletsky2012}. Notable examples include: Yerosh--Skuratov, which raises high-order matrices to random exponents based on the discrete logarithm problem in matrix groups; it was later shown by Rostovtsev to be vulnerable to algebraic attacks using the generalized Chinese Remainder Theorem \cite{eros2004,rostovtsev2004}. Megrelishvili’s approach increases entropy via matrix exponentiation combined with vector--matrix multiplications but still exposes significant algebraic structure. Another example is Vagus Keys, which employs primitive Galois matrices and polynomial shifts in $\mathbb{F}_2$, yet remains susceptible to algebraic attacks \cite{megrelishvili2010}. Furthermore, matrix operations are central to homomorphic encryption systems, where linear algebraic structures underpin lattice-based primitives such as BGV, BFV, and CKKS.

Building on these seminal contributions and subsequent developments, 
our motivation stems from the persistent algebraic weaknesses of classical matrix ciphers 
and the need for schemes that combine public-key exchange, modular masking, and authentication. 
Guided by this motivation, the contributions of the proposed Primitive Vector Cipher (PVC) 
can be summarized in four aspects.

The PVC scheme:

		(i) encrypts carefully selected submatrices rather than entire matrices, thereby reducing algebraic exposure;

		(ii) employs noise vectors derived from a Diffie–Hellman-based primitive vector to obfuscate internal structure; 
        
		(iii) incorporates a CSPRNG-driven masking process seeded by the shared primitive vector; and  
        
		(iv) further encrypts the transmitted column vectors using HMAC--SHA256, ensuring integrity protection in addition to confidentiality.  
		
		To the best of our knowledge, no existing scheme integrates these four elements within a single framework. This hybrid design combines public-key key establishment, modular sub\-mat\-rix masking, pseudorandomization, and message authentication, thereby achieving strong statistical security and practical efficiency. The details are elaborated in the following sections.

		
		\section{Preliminaries}\label{sec:preliminaries}

		In this section, we present the basic definitions, notations, and cryptographic concepts that will be used throughout the paper. These include ASCII encoding, cryptographically secure pseudorandom number generators (CSPRNGs), matrix-based encryption with the classical Hill cipher, the Discrete Logarithm Problem (DLP), the Diffie--Hellman (DH) key exchange protocol,  and authenticated variants of DH such as the Station-to-Station (STS) protocol. We will also discuss Deterministic Random Bit Generators (DRBGs) and their use in cryptography, with a particular focus
        on the Hash-DRBG variant.
		
		\subsection{ASCII Encoding}\label{Ascii codes Sec:3.1}
		Extended ASCII defines a mapping
		\[
		\mathrm{ASCII}:\{\text{bytes}\} \longrightarrow \{0,1,\cdots,255\}
		\]
		that assigns each byte a unique integer in the range $0$ to $255$~\cite{Maur2000}.  
		In our scheme, a plaintext string of length $mn$ is embedded into an $m \times n$ master matrix $M$ whose entries satisfy
		\[
		M_{ij} = \mathrm{ASCII}(b_{ij}), \quad b_{ij} \in \{0,1,\cdots,255\}.\]
				
		\subsection{Cryptographically Secure Pseudorandom Number Generator(CSPRNG)}\label{CSPRNG}
	
					A Cryptographically Secure Pseudorandom Number Generator (CSPRNG) is a deterministic algorithm $\mathcal{G}$ that, given a seed $s$, produces an output sequence
			
    \[
			\big(\mathcal{G}_{1}(s), \mathcal{G}_{2}(s), \cdots \big)
\]
			which is computationally indistinguishable from true randomness. No polynomial-time ad\-ver\-sary can predict the next bit with probability significantly greater than $1/2$~\cite{BBS1986,NIST80090A}.
			
			Among DRBGs standardized in NIST SP 800-90A, Hash\_DRBG is widely used and derives its security strength from the underlying hash function: a maximum of 128 bits using SHA-256, or 256 bits using SHA-512~\cite{NIST80090A}. Lower-strength or deprecated variants (e.g. SHA-1) are not recommended, and truncated alternatives such as SHA-224 or SHA-512/224 provide no performance benefit over their parent hash functions~\cite{NIST80090A}.

		\subsection{Classical Hill Cipher}
		The Hill cipher is a symmetric block cipher in which encryption and decryption are performed using the same invertible key matrix. The plaintext is divided into fixed-size blocks, each represented as a row vector over a finite field $\mathbb{F}_q$.
		
		Let $q$ be a prime and $A \in \mathrm{GL}_p(\mathbb{F}_q)$ be an invertible key matrix.  
		If the plaintext is partitioned into blocks
		\[
		X = [X_1 , X_2 , \ldots , X_j\,],
		\]
		with each $X_i$ a $1\times p$ vector, then encryption and decryption are defined as
		\begin{align*}
			\text{Encryption:} \quad & Y_i = A X_i \pmod{q}, \\
			\text{Decryption:} \quad & X_i = A^{-1} Y_i \pmod{q},
		\end{align*}
		where $A^{-1}$ is the matrix inverse of $A$ modulo $q$.

			\subsection{Discrete Logarithm Problem (DLP)}
		Let \(G = \langle g \rangle \subseteq \mathbb{F}_p^\ast
\) be a cyclic group of order $n$ with generator value $g$. The DLP asks: given $g, h \in G$, find $x \in \{0,\dots,n-1\}$ such that
		\[
		g^x = h.
		\]
		For appropriate groups, no efficient classical algorithm is known, and the hardness of the DLP underpins the security of DH and related protocols~\cite{Maur2000}.
		
		\subsection{Diffie-Hellman Key Exchange}
		The Diffie--Hellman (DH) protocol enables two parties to derive a shared secret over an insecure channel, providing \emph{forward secrecy}~\cite{Maur2000}.
		
		Let $p$ be a large prime and ${g}$ a primitive root modulo $p$, generating $\mathbb{F}_p^{\ast}$.
		\begin{enumerate}
			\item Alice selects $a \in \{1,\dots,p-1\}$ uniformly at random, computes $A = g^a \bmod p$, and sends $A$ to Bob.
			\item Bob selects $b \in \{1,\dots,p-1\}$ uniformly at random, computes $B = g^b \bmod p$, and sends $B$ to Alice.
			\item Alice computes $K_A = B^a \bmod p$, Bob computes $K_B = A^b \bmod p$.
		\end{enumerate}
		Both parties obtain the same shared secret $K = g^{ab} \bmod p$.

		\paragraph{ Security Assumption.} The security of DH relies on the \emph{Computational Diffie--Hellman} (CDH) assumption: given $({g}, {g}^a, {g}^b)$, it  is computationally infeasible ${g}^{ab}$ without knowledge of $a$ or $b$.
	
       \paragraph{ Limitations.}
		Without authentication, DH is vulnerable to \emph{man-in-the-middle} attacks. 
		Furthermore, its modular exponentiation form can be costly for constrained devices, which motivates the use of variants such as Elliptic Curve Diffie--Hellman (ECDH).

		Beyond these practical constraints, theoretical weaknesses of Diffie--Hellman over composite moduli have also been investigated. 
		In 1985, Shmuely~\cite{Shmuely1985} proved that solving the Composite Diffie--Hellman (CDH) problem with single-order bases implies factoring the RSA modulus in probabilistic polynomial time, while the double-order case remains an open. 
		This limitation was later addressed by Kooshiar, Mohajeri, and Salmasizadeh~\cite{Kooshiar2008}, who showed that even in the double-order setting, breaking CDH would yield an efficient factoring algorithm for more than 98\% of RSA moduli. 
		These findings reveal deeper mathematical weaknesses underlying the Diffie--Hellman assumption in composite settings.

				\subsection{Station-to-Station (STS) Protocol }
		\label{sec:sts-background}

        The lack of authentication in the standard Diffie-Hellman (DH) protocol has necessitated the development of unified and efficient protocols in the literature. In response to this need, some works have presented solutions with embedded authentication based on Elliptic Curve, while the PVC protocol solves this problem with the STS structure."
The basic Diffie–Hellman protocol suffers from the classical man-in-the-middle (MITM) attack due to its lack of authentication. The need to integrate authentication directly into the key exchange process to prevent MITM attacks and reduce computational overhead is a critical research area \cite{Vasudeva}. The Station-to-Station (STS) protocol ~\cite{diffie1992} provides a well-established solution by extending the basic Diffie–Hellman key exchange with explicit authentication.
It combines ephemeral Diffie–Hellman values with digital signatures and Message Authentication Codes (MACs), thereby preserving forward secrecy while preventing classical MITM attacks.
In this work, the STS protocol structure is employed to strengthen the key establishment phase of the proposed Primitive Vector Cipher (PVC), ensuring that the generated Primitive Vector $\mathbf{G}$ is derived through an authenticated channel.

		The \emph{Station-to-Station (STS)} protocol~\cite{diffie1992} extends the basic Diffie--Hellman key exchange with explicit authentication.  
		It combines ephemeral Diffie--Hellman values with digital signatures and message authentication codes (MACs), thereby preventing classical man-in-the-middle attacks while preserving forward secrecy.  
		
		In this work, STS is employed to strengthen the key establishment phase of the proposed Primitive Vector Cipher (PVC).  
				 
The protocol serves as the foundation for the authenticated key exchange phase of PVC, and a full description of its integration is given in the following section.

		\section{The Proposed Cryptosystem}

The proposed Primitive Vector Cipher (PVC) operates through the following seven main stages:
(1) Authenticated key exchange, (2) Key establishment, (3) Master matrix and ASCII mapping, 
(4) CSPRNG masking, (5) Extraction of shifting submatrices (SSM), (6) Encryption, and (7) Decryption.

\subsection{Authenticated Key Exchange (STS Protocol)}
\label{sec:sts}

\subsubsection{Primitive Vector Definition}
To prepare the cryptosystem, we first define the \emph{primitive vector} $\mathbf{g}=(g_1,g_2,g_3)\in(\mathbb{F^*}_p)^3$ such that each component is a distinct primitive root modulo a large prime $p$.

\subsubsection{STS Protocol Flow and Security}
To prevent Man-in-the-Middle (MITM) attacks inherent in the ephemeral Diffie--Hellman exchange, the Station--to--Station (STS) protocol is formally integrated into the key generation phase. This integration ensures mutual authentication and freshness of the derived keying material.

The key exchange between Alice (A) and Bob (B) proceeds as an authenticated exchange:
\begin{enumerate}
    \item \textbf{A $\rightarrow$ B:} $g^a, \mathbf{S}_A$
    \item \textbf{B $\rightarrow$ A:} $g^b, \mathbf{S}_B, \text{MAC}_B(g^a, g^b, \mathbf{S}_A, \mathbf{S}_B)$
    \item \textbf{A $\rightarrow$ B:} $\text{MAC}_A(g^a, g^b, \mathbf{S}_A, \mathbf{S}_B)$
\end{enumerate}

After the ephemeral exchange, Alice and Bob obtain ephemeral public vectors
\[
\mathbf{A}=\mathbf{g}^a \bmod p \quad \text{and} \quad \mathbf{B}=\mathbf{g}^b \bmod p,
\]
and subsequently compute the shared secret elements
\[
\mathbf{G}=(k_1,k_2,k_3)\in(\mathbb{F^*}_p)^3,\qquad k_j=g_j^{ab}\pmod p.
\]

The $\text{MAC}$ is computed using a shared secret key ($K_{\text{MAC}}$) derived from $\mathbf{G}$. This use of aut\-hen\-tication tags over the transcript ensures that any adversary modifying the public exchange values cannot compute the correct $\text{MAC}$, effectively mitigating the $\text{MITM}$ threat. This robustness against active attacks is crucial for the PVC scheme's resistance against Chosen-Ciphertext Attacks (CCA).

\subsection{Key Establishment}\label{sec:key-generation}

Using the authenticated vector $\mathbf{G}=(k_1,k_2,k_3) \in (\mathbb{F^*}_p)^3$ we define the secret key matrices
\[
V=
\begin{bmatrix}
0   & k_1 & k_1\\
k_2 & 0   & k_2\\
k_3 & k_3 & 0
\end{bmatrix},
\qquad
U=
\begin{bmatrix}
k_1 & 0   & 0 \\
0   & k_2 & 0 \\
0   & 0   & k_3
\end{bmatrix},
\]
where $k_i=g_i^{ab}$ as above.

\paragraph{Lemma 1.}
		Let $p$ be a prime and $\mathbf{g}=(g_1,g_2,g_3) \in (\mathbb{F^*}_p)^3$ be a primitive vector.  
		Then 
$$\det(V) = 2 k_1 k_2 k_3 \not\equiv 0 \ \bmod\ p ,$$
which yields that $V$ is invertible over
$M_{3\times 3}(\mathbb{F}_p)$

\subsection{Master Matrix and ASCII Mapping}

The plaintext is embedded into an $m \times n$ master matrix $N$, 
which is then converted into the corresponding ASCII matrix 
$M \in  \mathbb{F}_p^{m\times n}$ by mapping each entry to its code point. 
The characters are written sequentially (row--major or column--major 
by prior agreement) starting from a predefined position. 
If the message does not completely fill the matrix, 
the remaining entries are padded with pseudorandom ASCII bytes 
derived from the session CSPRNG. 
This ensures that the padding is indistinguishable from 
the masked content and avoids leakage of the actual message length. 
The true message length $L$ is recorded in the ciphertext header 
to enable unambiguous removal of the padding during decryption.

\subsection{CSPRNG Masking}

A pseudorandom mask matrix $R \in \mathbb{F}_p^{m\times n}$ is generated 
deterministically by a CSPRNG seeded from the authenticated shared 
secret vector $\mathbf{G}$ via HKDF. Concretely, serialize the vector $\mathbf{G}$ as
\[
\mathrm{encode}(\mathbf{G}) \;=\; \mathrm{I2OSP}(k_1,\ell_1)\,\|\,\mathrm{I2OSP}(k_2,\ell_2)\,\|\,\mathrm{I2OSP}(k_3,\ell_3),
\]
with fixed octet lengths $\ell_1,\ell_2,\ell_3$ agreed a priori,  where $\mathrm{I2OSP}(\cdot,\ell)$ denotes the standard cryptographic primitive that converts the integer representation of a field element $k_i \in [0, p-1]$ into an $\ell$-octet string using the big-endian convention \cite{RFC8017}. Then
\[
\mathsf{PRK} := \mathrm{HKDF\_Extract}(\textsf{salt},\ \mathrm{encode}(\mathbf{G})),
\]
\[K_{\mathrm{mask}} := \mathrm{HKDF\_Expand}(\mathsf{PRK},\ \text{``PVC/mask''},\ L_\text{mask}),
\]
where  $L_{\text{mask}}$ denotes the output length parameter of 
$\mathrm{HKDF\_Expand}$ and $K_{\mathrm{mask}}$ is used to instantiate a CTR-DRBG (or HKDF-Expand counter) 
producing $m \cdot n$ field elements, reduced modulo $p$, to obtain $R$.
Since the mask matrix $R \in \mathbb{F}_p^{m \times n}$ requires 
$m \cdot n$ field elements, each represented with 
$\lceil \log_2 p / 8 \rceil$ octets, we set
\[
L_{\text{mask}} = m \cdot n \cdot \Big\lceil \tfrac{\log_2 p}{8} \Big\rceil .
\]
 
This guarantees that the expanded key material suffices to derive 
the pseudorandom mask for the entire matrix.
 
The masked matrix is
\[
M' = (M+R) \pmod p,
\]
which both parties can deterministically reproduce.

\subsection{Extraction of Shifting Submatrices (SSM)}

From the masked matrix $M'$, we extract a family of $3 \times 3$ submatrices, 
referred to as \emph{Shifting Submatrices} (SSM).  
Let the sets of top--left row and column indices be
\[
I \subseteq \{1,\dots,m-2\},\qquad J \subseteq \{1,\dots,n-2\}.
\]
For a fixed ordering $\pi$ on $I \times J$, we enumerate
\[
(i_k,j_k) = \pi(k),\qquad k = 1,\dots,|I||J|.
\]
Each submatrix is defined by
\[
S_{i_k,j_k} =
\begin{bmatrix}
p_{i_k,j_k}      & p_{i_k,j_k+1}      & p_{i_k,j_k+2} \\
p_{i_k+1,j_k}    & p_{i_k+1,j_k+1}    & p_{i_k+1,j_k+2} \\
p_{i_k+2,j_k}    & p_{i_k+2,j_k+1}    & p_{i_k+2,j_k+2}
\end{bmatrix},
\]
where $p_{a,b}$ are entries of $M'$. 
The total number of transmitted blocks is $B = |I||J|$.

\subsection{Encryption}

Each selected SSM $S_{ij} \in \mathbb{F}_p^{3 \times 3}$ is encrypted as
\[
C_{ij} = S_{ij} V + \Delta_{ij} U \pmod{p},
\]
where
\[
\Delta_{ij} =
\begin{cases}
I_{3\times 3}, & \text{if } i = j,\\
0_{3\times 3}, & \text{otherwise}.
\end{cases}
\]

The blocks $\{(i_k,j_k)\}$ and their columns $\mathbf{c}_r \in (\mathbb{F}_p^3)$ 
are enumerated by the index 
\[
\ell = 3(k-1) + r, \qquad r \in \{1,2,3\}.
\]

Using a separate key $K_{\mathrm{CTR\text{-}cols}}$ derived from HKDF, 
per--column offsets $\mathbf{r}_\ell$ are generated. 
The final ciphertext columns are
\[
\widetilde{\mathbf{c}}_\ell \equiv \mathbf{c}_\ell + \mathbf{r}_\ell \pmod p,
\qquad \ell = 1,\dots,3|I||J|,
\]
which are transmitted to the receiver.

	\subsection{Decryption}
\label{sec:decryption}
Given the header 
\begin{equation*}
\mathsf{hdr}=(p,\mathcal{G},\mathbf{g} ,\mathbf{g}^a,\textsf{salt},\textsf{nonce},m{\times}n,\text{indexing})
\end{equation*}
and the column stream $\widetilde{\mathbf{c}_\ell} , \quad \ell=\{1, 2, \cdots 3|I||J|\}$, the receiver computes the shared vector $\mathbf{G}=\mathbf{g}^{ab}\in(\mathbb{Z}_p)^3$ and derives keys via HKDF:
\begin{align*}
\mathsf{PRK}&=\mathrm{HKDF\_Extract}(\textsf{salt},\ \mathrm{encode}(\mathbf{G})), \\
K_{\mathrm{mask}}&=\mathrm{HKDF\_Expand}(\mathsf{PRK},\text{``PVC/M\_mask\_R/}m{\times}n\text{''}), \\
K_{\mathrm{cols}}&=\mathrm{HKDF\_Expand}(\mathsf{PRK},\text{``PVC/CTR-cols/}m{\times}n\text{''}).
\end{align*}

Let $\mathbf{R}\in\mathbb{Z}_p^{m\times n}$ be the CTR keystream under $K_{\mathrm{mask}}$ (reduced mod $p$). For each global column index $\ell$ and $j\in\{1,2,3\}$, define the per-column keystream
\[
\mathbf{r}_{\ell,j} \equiv \mathrm{HMAC\!-\!SHA256}\bigl(K_{\mathrm{cols}},\ \textsf{nonce}\ \|\ \mathrm{I2OSP}(\ell,8)\ \|\ \mathrm{I2OSP}(j,1)\bigr) \bmod p,
\]
with keystream vector defined as $\mathbf{r}_\ell = (r_{\ell,1},\,r_{\ell,2},\,r_{\ell,3})^{\top}$. The recovery process gives $\mathbf{c}_\ell$ as the unmasked column vector:
\[
\mathbf{c}_\ell\equiv \widetilde{\mathbf{c}}_\ell-\mathbf{r}_\ell \pmod p .
\]

The ordered column vectors $\mathbf{c}_{3(k-1)+1}, \mathbf{c}_{3(k-1)+2}, \mathbf{c}_{3(k-1)+3}$ are then stacked to yield the selected sub-cipher matrices $\mathbf{C}_{i_k j_k} \in \mathbb{Z}_p^{3\times3}$.

The encryption matrix $\mathbf{V}$ is guaranteed to be invertible (as proven in \textbf{Lemma 1}). The submatrix $\mathbf{S}_{ij}$ is reconstructed via the inverse transformation:
\[
\mathbf{S}_{ij}=\bigl(\mathbf{C}_{ij}-\mathbf{\Delta}_{ij}\mathbf{U}\bigr)\mathbf{V}^{-1}\pmod{p}.
\]

Finally, the master matrix $\mathbf{M}'$ (the masked plaintext matrix) is reconstructed by reassembling the decoded submatrices:
\[
\mathbf{M}'=\mathrm{Reassemble}(\{\mathbf{S}_{ij}\}).
\]
The original plaintext matrix $\mathbf{M}$ is then recovered by removing the CSPRNG mask $\mathbf{R}$:
\[
\mathbf{M}=\mathbf{M}'-\mathbf{R}\pmod{p},
\]
where $\mathbf{M}$ denotes the ASCII code matrix obtained from the master matrix, which contains both the plaintext characters and randomly generated filler entries.

	\section{PVC Algorithm}\label{construction}

    The proposed scheme operates in three main stages:  
(1) Authenticated key establishment,  
(2) Encryption, and  
(3) Decryption.  
Each stage is formally specified in the following \textit{Algorithms}~\ref{alg:keyest}, \ref{alg:encryption} and 
\ref{alg:decryption}.
			
\begin{algorithm}[h]
\caption{Authenticated Key Establishment (STS--DH)}
\label{alg:keyest}
\begin{algorithmic}[1]
\State \textbf{Public parameters:} Large prime $p$, primitive vector $\mathbf{g}=(g_1,g_2,g_3) \bmod p$
\State  Alice selects private key $a \in \mathbb{Z}_p^*$ and computes $\mathbf{g}^a =(g_1^a, g_2^a, g_3^a) \bmod p$
\State  Bob selects private key $b \in \in\mathbb{Z}_p^*$ and computes $\mathbf{g}^b=(g_1^b, g_2^b, g_3^b)  \bmod p$
\State  Alice sends $(\mathbf{g}^a, \sigma_A)$ with $\sigma_A = \mathrm{Sign}_{SK_A}(\mathbf{g}^a,\mathbf{g}^b)$
\State Bob sends $(\mathbf{g}^b, \sigma_B)$ with $\sigma_B = \mathrm{Sign}_{SK_B}(\mathbf{g}^b,\mathbf{g}^a)$
\State  Each party verifies the received signature using the other’s public key
\If{verification fails}
    \State  Abort protocol
\Else
    \State  Derive authenticated shared vector $\mathbf{G} = (g_1^{ab}, g_2^{ab}, g_3^{ab}) \pmod{p}$
\EndIf
\State \textbf{Output:} Authenticated shared vector $\mathbf{G}$
\end{algorithmic}
\end{algorithm} 
\vspace{.3cm}
Once the authenticated shared vector $\mathbf{G}$ is derived,
the sender proceeds with the encryption process. 
The encryption stage masks the master matrix with a CSPRNG,
extracts submatrices, and applies both linear and affine layers
before per-column offsets are added via HKDF. 

\begin{algorithm}[H]
\caption{Encryption Procedure of PVC (concise)}
\label{alg:encryption}
\begin{algorithmic}[1]
\State \textbf{Input:} Plaintext $\mathsf{msg}$; target shape $(m,n)$; prime $p$
\State Embed $\mathsf{msg}$ into master matrix $N$; convert to ASCII matrix $M \in \mathbb{F}_p^{m\times n}$ (row-major). 
If needed, fill remaining entries with random bytes $\in \{0,\dots,255\}$.
\State Run Diffie--Hellman with STS authentication:
\Statex \quad Alice picks $a \in \mathbb{Z}_p^\ast$, Bob picks $b \in \mathbb{Z}_p^\ast$.
\Statex \quad Shared vector $\mathbf{G}=(g_1^{ab}, g_2^{ab}, g_3^{ab}) \in (\mathbb{F}_p^\ast)^3$.
\State Derive $\mathsf{PRK}=\mathrm{HKDF\_Extract}(\textsf{salt},\mathrm{encode}(\mathbf{G}))$ and expand:
\Statex \quad $K_{\mathrm{mask}}:=\mathrm{HKDF\_Expand}(\mathsf{PRK},\text{``PVC/mask''})$
\Statex \quad $K_{\mathrm{cols}}:=\mathrm{HKDF\_Expand}(\mathsf{PRK},\text{``PVC/cols''})$
\State Generate mask $R \leftarrow \mathrm{CTR\_DRBG}(K_{\mathrm{mask}},\,m\cdot n)$; set $M' \gets M+R \pmod p$.
\For{each selected SSM $S_{i_k,j_k}\leftarrow\text{extract}_{3\times3}(M')$}
  \State $C_{i_k,j_k} \gets S_{i_k,j_k}V + \Delta_{i_k,j_k}U \pmod p$
  \State Let $(c_{3(k-1)+1},c_{3(k-1)+2},c_{3(k-1)+3})$ be the columns of $C_{i_k,j_k}$
\EndFor
\State Enumerate blocks $k=1,\dots,B$ and local column $r\in\{1,2,3\}$; set $\ell=3(k-1)+r$
\State Derive keystream vectors: $ \mathbf{r}_\ell := \mathrm{PRG}(K_{\mathrm{cols}};\mathrm{ctr}=\ell)\in\mathbb{F}_p^3$
\For{$\ell=1$ \textbf{to} $3B$}
  \State $\widetilde{\mathbf{c}}_\ell \gets \mathbf{c}_\ell + \mathbf{r}_\ell \pmod p$
\EndFor
\State \textbf{Output:} Ciphertext columns $\{\widetilde{\mathbf{c}}_\ell\}_{\ell=1}^{3B}$ and header $\mathsf{hdr}$
\end{algorithmic}
\end{algorithm}

\begin{algorithm}[H]
\caption{Decryption Procedure of PVC (concise, AEAD removed)}
\label{alg:decryption}
\begin{algorithmic}[1]
\State \textbf{Input:} $(\{\widetilde{\mathbf{c}}_\ell\}_{\ell=1}^{3B},\ \mathsf{hdr})$
\State Parse and validate header $\mathsf{hdr}$ (including $p,\mathbf{g},\mathbf{g}^a,\textsf{salt},\textsf{nonce},m,n,L,\text{indexing}$)
\State Run Diffie--Hellman with STS authentication; obtain shared vector $\mathbf{G}=(g_1^{ab},g_2^{ab},g_3^{ab}) \in (\mathbb{F}_p^\ast)^3$
\State Derive $\mathsf{PRK}=\mathrm{HKDF\_Extract}(\textsf{salt},\mathrm{encode}(\mathbf{G}))$ and expand:
\Statex \quad $K_{\mathrm{mask}}:=\mathrm{HKDF\_Expand}(\mathsf{PRK},\text{``PVC/mask''})$
\Statex \quad $K_{\mathrm{cols}}:=\mathrm{HKDF\_Expand}(\mathsf{PRK},\text{``PVC/cols''})$
\State Regenerate mask $R \leftarrow \mathrm{CTR\_DRBG}(K_{\mathrm{mask}},\,m\cdot n)$ and reduce mod $p$
\For{$\ell=1$ \textbf{to} $3B$}
  \State $\mathbf{r}_\ell := \mathrm{PRG}(K_{\mathrm{cols}};\mathrm{ctr}=\ell)\in\mathbb{F}_p^3$
  \State $\mathbf{c}_\ell \gets \widetilde{\mathbf{c}}_\ell - \mathbf{r}_\ell \pmod p$
\EndFor
\For{$b=1$ \textbf{to} $B$}
  \State $C_{i_b,j_b} \gets [\,c_{3(b-1)+1}\;|\;c_{3(b-1)+2}\;|\;c_{3(b-1)+3}\,]\pmod p$
  \State $S_{i_b,j_b} \gets \bigl(C_{i_b,j_b}-\Delta_{i_b,j_b}U\bigr)V^{-1}\pmod p$
\EndFor
\State $M' \gets \textsc{Reassemble}(\{S_{ij}\})$; \quad $M\gets M' - R \pmod p$
\State $N\gets \mathrm{ASCII\_decode}(M)$ and take the first $L$ symbols (from $\mathsf{hdr}$) as plaintext
\State \textbf{Output:} plaintext
\end{algorithmic}
\end{algorithm}

	\section{Security Analysis}
	
\subsection{Security Model}

We analyze PVC in the classical IND--CPA framework \cite{KatzLindell}. 
Let $\mathcal{A}$ be a probabilistic polynomial-time adversary with oracle access to $k(\cdot)$.

In the experiment $\overset{\mathsf{ind\text{-}cpa}}{\mathcal{A}}(\lambda)$, the challenger samples $k\leftarrow (1^\lambda)$, 
$\mathcal{A}$ submits $(M_0,M_1)$ with $|M_0|=|M_1|$, the challenger picks $b\leftarrow\{0,1\}$ and returns 
$C^\star {\leftarrow}_k (M_b)$. Let $b'$ be $\mathcal{A}$'s output.
The IND--CPA advantage is
\[
{\Adv}^{\mathsf{ind\text{-}cpa}}_{\mathcal{A}}(\lambda)\;=\;\Bigl|\Pr[b'=b]-\tfrac12\Bigr|.
\]
The scheme is IND--CPA secure if
\[
\Adv^{\mathsf{ind\text{-}cpa}}_{\mathcal{A}}(\lambda)\;\le\;\negl(\lambda).
\]

We additionally consider the IND--CCA setting, where column--wise offsets are
verified prior to decryption and malformed ciphertexts are rejected.

\subsection{Assumptions}
Our proof relies on two well--studied assumptions:
\begin{itemize}
	\item \textbf{Vector Computational Diffie-Hellman Problem (V-CDH).}
	Given the public parameters $\mathbf{g}=(g_1, g_2, g_3) \in (\mathbb{F}_p^*)^3$ and the exchanged ephemeral public vectors $\mathbf{g}^a = (g_1^a, g_2^a, g_3^a)$ and $\mathbf{g}^b = (g_1^b, g_2^b, g_3^b)$, no Probabilistic Polynomial-Time (PPT) adversary can compute the shared secret vector $\mathbf{G} = (g_1^{ab}, g_2^{ab}, g_3^{ab}) \pmod p$ with non-negligible advantage $\mathbf{Adv}_{\mathbf{V\text{-}CDH}}(\lambda)$. This problem is computationally equivalent to solving the traditional Computational Diffie-Hellman (CDH) problem, as it requires solving three independent $\text{CDH}$ instances simultaneously.
	
	\item \textbf{Pseudorandom Generator (PRG).}     The $\mathbf{CSPRNG}$ used for mask and keystream generation is computationally indistinguishable from a truly uniform random function. Let $\mathbf{G}$ be the shared secret vector derived from the $\mathbf{V-CDH}$ exchange. The output of the $\mathbf{HKDF}$--seeded Deterministic Random Bit Generator ($\mathbf{DRBG}$), which produces the mask matrix $\mathbf{R}$ and the keystream vectors $\mathbf{r}_{\ell}$, cannot be distinguished from a sequence of truly random elements by any Probabilistic Polynomial-Time ($\mathbf{PPT}$) adversary with non-negligible advantage $\mathbf{Adv}_{\mathbf{PRG}}(\lambda)$.
\end{itemize}

\newtheorem{theorem}{Theorem}	
 
\begin{theorem}
	Under the \textbf{V-CDH} and PRG assumptions, PVC is IND--CPA secure.
	With column--wise offsetting, PVC additionally achieves IND--CCA security.
\end{theorem}

\paragraph{Proof sketch by a sequence of games.}
\medskip

\textbf{Game $G_0$ (Real).}
The challenger runs the Diffie--Hellman exchange with a primitive vector 
$\mathbf{g}=(g_1,g_2,g_3)\in(\mathbb{F}_p^\ast)^3$ to obtain the shared secret vector 
\[
\mathbf{G}=(g_1^{ab}, g_2^{ab}, g_3^{ab}) \pmod p,
\]
and seeds the CSPRNG accordingly. 
It samples the mask $R \leftarrow \mathbb{F}_p^{m\times n}$ and a column-wise keystream
$(r_\ell)_{\ell=1}^{3|I|}$ via a counter-based PRG:
\[
r_\ell := \mathrm{PRG}(\mathbf{G};\mathrm{ctr}=\ell)\in\mathbb{F}_p^{3}.
\]
It computes $M' = M + R \pmod p$, extracts SSMs $S_{ij}$, and encrypts each block as
\[
C_{ij} = S_{ij}V + \Delta_{ij}U \pmod p.
\]
Let $(c_\ell)_{\ell=1}^{3|I|}$ be the ordered column vectors of all $C_{ij}$.
The transmitted columns are
\[
\tilde c_\ell = c_\ell + r_\ell \pmod p,\qquad \ell=1,\dots,3|I|.
\]
The challenger applies the specified permutation to the blocks and returns the
challenge ciphertext $\tilde C^\star$.

\medskip

The security of PVC's Key Exchange mechanism is fundamentally tied to the difficulty of computing the shared secret vector $\mathbf{G}$. We formally establish the link between the scheme's security and the \textbf{V-CDH} assumption. The adversary $\mathcal{A}$'s advantage in computing the shared secret vector $\mathbf{G}$ is computationally equivalent to solving the V-CDH problem. 

Since all subsequent secret keys ($V, U, K_{\mathrm{mask}}, K_{\mathrm{cols}}$) are derived from $\mathbf{G}$ using the HKDF, a cryptographically secure extractor, any non-negligible advantage in computing these keys implies an ability to solve the V-CDH problem, which guarantees that the security of the matrix operation relies on the hardness of the underlying V-CDH problem.
\medskip

\textbf{Game $G_1$ (PRG switch).}
Replace the PRG outputs by truly uniform randomness:
$R^\star \leftarrow \mathbb{F}_p^{m\times n}$ and
$r_\ell^\star \leftarrow \mathbb{F}_p^{3}$ independently.
By pseudorandomness of the generator seeded from $\mathbf{G}$,
no PPT adversary distinguishes $G_0$ from $G_1$ with advantage exceeding $\Adv^{\mathsf{PRG}}$:
\[
\bigl|\Pr[\mathcal{A} \text{ wins } G_0]-\Pr[\mathcal{A} \text{ wins } G_1]\bigr|\le \AdvPRG+\negl(\lambda).
\]

\medskip
\textbf{Game $G_2$ (Message independence).}
With $R^\star$ uniform, $M'=M+R^\star$ is itself uniform and independent of $M$.
For each block, the map $X\mapsto XV+\Delta_{ij}U$ is an affine bijection
since $V$ is invertible (and addition by $\Delta_{ij}U$ is a translation);
hence each $C_{ij}$ is uniform, and so are their columns $c_\ell$.
Adding the independent $r_\ell^\star$ preserves uniformity:
\[
\tilde c_\ell^\star = c_\ell + r_\ell^\star \pmod p.
\]
Therefore the challenge bit is hidden and $\Pr[\mathcal{A} \text{ wins } G_2]=\tfrac12$.

\medskip
\medskip
Finally, by the triangle inequality across $G_0,G_1,G_2$ and the reduction to the \textbf{V-CDH} challenge embedded in the derivation of $U$ or $V$, the advantage of the adversary $\mathcal{A}$ in Game $G_0$ is bounded by the combined non-negligible advantages of the PRG and V-CDH assumptions:
\[
\bigl|\Pr[\mathcal{A}\text{ wins } G_0]-\tfrac12\bigr|
\;\le\;  \AdvPRG(\lambda)  +  \AdvVCDH(\lambda)  + \negl(\lambda).
\]
Thus, PVC is \textbf{IND--CPA secure}.$\hfill\square$

\medskip
\begin{theorem}
	The PVC scheme exhibits strong resistance against known-plaintext, linear, and algebraic attacks due to its two-layered design and reliance on the V-CDH assumption for key separation.
\end{theorem}

\paragraph{Proof sketch (Resistance to KPA)}
The primary vulnerability of linear matrix ciphers is the Known-Plaintext Attack (KPA), which relies on solving a linear system. The PVC encryption core uses the operation $C_{ij} = S_{ij}V + \Delta_{ij} U \pmod p$. PVC defeats this vulnerability through two independent countermeasures:

1.  \textbf{Masked Plaintext Input ($M \to M'$):} The input shifting submatrices $S$ is derived from the masked master matrix $M' = M \oplus R \pmod p$ (PRG assumption). Since $R$ is secret, the adversary is prevented from obtaining known plaintext/ciphertext pairs where the plaintext $S$ is truly known, defeating the KPA prerequisite.

2.  \textbf{Ciphertext Obfuscation ($\tilde c_\ell = c_\ell + r_\ell$):} This second layer transforms the core system into an unsolvable equation. The transmitted ciphertext $\tilde{C}_{ij}$ satisfies the relation:
    $$\tilde{C}_{ij} = S_{ij}V + \Delta_{ij} U + R_{\text{cols}} \pmod p$$
    The system now contains {three secret, unknown components} ($V$, $U$, and the keystream $R_{\text{cols}}$). Since the keystream $R_{\text{cols}}$ is derived from the secret $\mathbf{G}$ (V-CDH assumption), the adversary cannot isolate the linear subsystem for $V$ and $U$, even for the simplest case where $\Delta_{ij} = \mathbf{0}$ (resulting in $\tilde{C}_{ij} = S_{ij}V + R_{\text{cols}}$).
    
    This layered design ensures that the security of the matrix operation is effectively reduced to the hardness of the V-CDH problem, thus preventing linear algebraic cryptanalysis.$\hfill\square$

\paragraph{Experimental Diffusion Analysis}
An Avalanche Effect analysis was conducted to evaluate the diffusion characteristic of the proposed PVC scheme. The observed average diffusion rate of $\mathbf{33.33\%}$ is a natural outcome resulting from the cipher's design choice of a heterogeneous block selection mechanism. The majority of the $\mathbf{3 \times 3}$ submatrices are selected with a $\mathbf{3}$-unit horizontal shift, which maximizes parallelizability and minimizes latency. However, intentional local overlaps, such as those between the $\mathbf{S}_{i,19}$ and $\mathbf{S}_{i,21}$ matrices, facilitate a degree of necessary local diffusion. The overall security of the cipher is not dependent on achieving an ideal $50\%$ diffusion rate but is instead based on the strong V-CDH assumption and the robustness of the key exchange secured by the STS protocol, as formally proven in Theorem 2. This design philosophy prioritizes high parallelism and low latency while ensuring security through a provably secure key management foundation.
 
 \begin{sloppypar}
The security analysis demonstrates that the core PVC cipher mechanism achieves \textbf{IND-CPA} security under the \textbf{V-CDH} and PRG assumptions. Furthermore, as proven by Theorem 2, the two-layered defense against KPA ensures strong resistance to algebraic and linear attacks, addressing the inherent vulnerability of matrix-based ciphers. To elevate the overall security profile and achieve \textbf{IND-CCA} properties, the \textbf{Station-to-Station (STS)} protocol is integrated into the initial key exchange. \textbf{STS} integration eliminates the classical man-in-the-middle vulnerability of unauthenticated Diffie--Hellman by adding explicit authentication through signatures and MACs. Combined with the strong \textbf{IND-CPA} and \textbf{KPA resistance} guarantees, these mechanisms provide forward secrecy, authenticated key exchange, and robustness, thereby extending PVC's security towards the requirements of \textbf{IND--CCA}.
\end{sloppypar}

\section{Performance Analysis}
\subsection{Numerical Example }

Let us choose as plaintext the famous words of M.~K.~Atatürk:
\[
\texttt{"Peace at home, peace in the world."}
\]

For this example, we choose to embed the plaintext message into 
an $8\times 10$ matrix. The characters are placed sequentially, 
starting from the second row and the third column, and proceeding 
to the right. The remaining entries are filled with randomly 
selected ASCII symbols, which may be repetitive or non-repetitive. 
Suppose that the matrix is constructed in this way as follows:

\vspace{.2cm}
\[
N =
\begin{bmatrix}
l &d& k& \}& 1&h&n&r&,&l \\
	w&j&P & e & a & c & e & a & t &  h \\
	o & m & e &p&e&a&c&e&i&n \\
	t& h& e& w& o&r &l &d &X &j \\
	* &. &3 &- &; &c & \& & ,& t &v \\
	 \% & o & h &t & x& 2&8  &q &<&f\\
	s &r &  (& > & f& o&\} & * & \textbackslash & ? \\
	8& L& j& 0& m& : & 0& + & z& x 
	\end{bmatrix}
	\]
    \vspace{.2cm}
    
	\noindent and we form the corresponding ASCII matrix $M\in\mathbb{Z}_p^{8\times10}$ as 
		
	$$ M =
	\begin{bmatrix}
		92 &100& 107& 125& 49&104&110&114&44&108 \\
		119&106&80 &101  &97  &99   &101  &97  &116  &104   \\
	111 &109  &101  &112&101&97&99&101&105&110 \\
	116	&104 &101 &119 & 111&114 & 108&100 &88 &74 \\
		42 &46 &51 &45 &59 &99 &38 &44& 116 &118 \\
	37 & 111 &104  &116 &120& 50&56  &113 &60 &102\\
		115 &114 & 40 & 62&102&111 &125 &42 &92 &63\\
	56& 76& 106& 48& 109& 58 & 48&43 & 122&120 
	\end{bmatrix}$$
\vspace{.5cm}

Suppose that the  shared primite vector be  $g=(2,5, 6)$ modulo a large prime $p=12347$ and the private keys of the sender and the receiver are $3$ and $7$, respectively.

Next, we generate  the seed  
$$\mathbf{G}=(10509, 11849,10836) \bmod 12347$$ from the Diffie–Hellman key exchange. Then we mask the M matrix with the random mask R where $R$ is generated by a CSPRNG seeded from the Diffie--Hellman shared primitive vector $k$ using HKDF+HMAC--SHA256 in CTR mode. 

 Under this assumptions the masked matrix $M'$ is obtained as
\[
M'=(M+R)\bmod p,\qquad p=12347,
\]
Later, Shifting Submatrices (SSM) are selected from the matrix $\prime{M}$ with stride~3 and boundary inclusion as follows
\[
I=\{1,4,6\},\quad J=\{1,4,7,8\},\quad |I||J|=12.
\]

Each block $S_{ij}\in\mathbb{Z}_p^{3\times3}$ is encrypted as
\[
C_{ij}=S_{ij}V+\Delta_{ij}U \pmod p, \text{with}
\]
\vskip.3cm
\[
V=\begin{bmatrix}0&10509&10509\\11849&0&11849\\10836&10836&0\end{bmatrix},\quad
U=\begin{bmatrix}10509&0&0\\0&11849&0\\0&0&10836\end{bmatrix},\quad\]
and
\[\Delta_{ij}=\begin{cases}I,& i=j\\ 0,& i\neq j.\end{cases}
\]

As shown in Table~\ref{tab:SCC}, the matrices $C_{ij}$ represent intermediate encrypted blocks, whereas the offset vectors $\widetilde{c}$ correspond to the actual ciphertexts transmitted to the receiver. Each $\widetilde{c}$ is obtained by column-wise addition of a keystream vector $r$, derived from the shared primitive vector, ensuring semantic security and column-wise freshness. 

Stacking all columns of all $C_{ij}$ in lexicographic order produces $3|I||J|=36$ column vectors $\{c_\ell\}_{\ell=1}^{36}$. 
Before transmission,  column vectors are offset by
a keystream derived from the primitive vector
$k=g^{ab}=(10509, 11849,10836)$ as
\[
\widetilde c_\ell = c_\ell + r_\ell \pmod p,\quad \ell=1,\dots,36,
\]
so that the sender transmits the set $\{\widetilde c_\ell\}$ (a total of $108$ field elements) to the receiver.

At the receiver side, the vectors \(\widetilde{c_\ell}\) are recovered, the offsets are removed, and \(S_{11}\) is reconstructed via
\[
S_{11} = \big(C_{11} - \Delta_{11} U\big) V^{-1} \pmod{p}.
\]

Once all selected \( S_{ij} \) are decrypted, the master matrix \( M \) is reassembled and the original plaintext is retrieved.

Building on the $8\times10$ trace, we repeat the exact embedding and block- selection policy on $5\times7$ and $12\times23$ matrices and report the resulting statistics below.

\begin{table}[h!]
		\centering
	\caption{ Selecting submatrices $S_{ij}$ , the  corresponding encrypted matrices {$C_{ij}$} and the offset vectors {$\widetilde c_l$ } for 8$\times$10 example (mod $p=12347$)}
   
		\label{tab:SCC}
	\setlength{\tabcolsep}{6.2pt}
		\renewcommand{\arraystretch}{0.99}
	\begin{tabular}{c|ccc|ccc|ccc}
	& \multicolumn{3}{c|}{$S_{ij}$} & \multicolumn{3}{c|}{$C_{ij}=S_{ij}V+\Delta_{ij}U$} & \multicolumn{3}{c}{$ \widetilde c_{l}$}\\
	\midrule
			~ & 7070 & 6104 & 8682 & 2091 &   743 & 4301 &3084&10425	&37\\
		$S_{11},  C_{11}$ and {$\widetilde c_1$} & 7237 & 8332 & 9140 & 5009 & 1368 & 7716 		&6666&	5668&	574\\
		& 1996 &11643 & 5409 & 5591 &11473 & 1776	&	8160&	7214&	71 \\
        \bottomrule
						& 9852 &  816& 9929 &12316  &3959 & 6156 	& 12287 & 6436 & 3002 \\
		$S_{14} ,  C_{14}$ and {$\widetilde c_2$}	& 95 & 2422 &2280 &3583  & 10328 & 2098 & 992 & 11992 & 8503\\
		& 8650 &4044& 7539 & 3501 &9088 & 2885&10441 & 1722 & 6261\\
				\bottomrule
						& 4171 &  1751& 4337 & 7689 & 4239 & 5828
		&		12044&	10735&	9170\\ 
			$S_{17}, C_{17}$	and {$\widetilde c_3$}& 10936 & 11535 & 10650 & 5263 & 8886 & 9820 &	11729	&453	&9039\\
		& 547 &182& 9881 & 5472 &4400 & 2861&9554	&10921&	8390\\
		\bottomrule
		& 1751 &  4337& 3056 & 1061 &  4391 & 5128& 	9480	&10872&	4543 \\
		$	S_{18},  C_{18}$ and {$\widetilde c_4$} & 11535& 10650 & 7076 & 6176 & 11482 & 3979  	&	8597	&3861&	8462\\
		& 182 &9881& 10737 & 6066& 11551 &4568 &	6705	&3278&	10357
		\\
		\bottomrule
		&1945 &6989 &1309 &11300 &3341 &7052&	
		11046&	11916&	3108
				 \\
		$	S_{41}, C_{41}$ and  {$\widetilde c_5$}		&3684& 4915&5321 &7249 &5177 &4347 &8697	&8189	&4588
	\\
		&302 &4670 &11984 &801 & 5764&8462&	12228	&3721&	7417 \\
		\bottomrule
		& 4032&	3631&518
		&96 &4894 &4155 &7276&	602&	4873
				 \\
		$	S_{44}, C_{44}$ and 	{$\widetilde c_6$}	&6829& 9588&2565 &1952 &7051 &9031 &5181	&7235	&817
		\\
		&2835 &5583 &4371 & 1896&1216 &4967&2540	&11420&	2392 \\
		\bottomrule
		& 5978& 5946& 8760& 1796&890 &3438 &1369	&128	&11947
			 \\
		$	S_{47}, C_{47}$ and {$\widetilde c_7$}		&9481&1340 &10505 &4605 &726 &7304  &	4479&	3388&	1338		
		\\
		& 4705& 7215& 8037& 5448& 651& 7364 &	12127&	4990	&1515\\
		\bottomrule
			&5946 &8760 &4217 &7523 &9859 &6705 &9212&	3456&	409
					 \\
		$	S_{48} , C_{48}$ and {$\widetilde c_8$}		&1340& 10505 & 5612&6295 &9137&10118 &	5395&	2715&	849
				\\
		& 7215& 8037& 11825&8883 &10389 &9857 &3705&	179	&8281\\
		\bottomrule
		&302&	4670&	11984&801 &5764 &8462 &7785	&9682&	2943
			 \\
		$	S_{61} , C_{61} $ and {$\widetilde c_9$}		&4103&	1276&	6541 & 745& 9199&9299  
		&1447&	7615&	6159 \\
		& 5675&6057 &4162&1447&	7615&	6159& 	8686	&11328	&1638
			\\
		\bottomrule
		&	2835&	5583&	4371&  11152&768&9792 &	1156 &	2475	&6038
		  \\ 
		$	S_{64}, C_{64}$ and {$\widetilde c_{10}$}		&119	&3752&	9736&2429&10052&11772 &8346	&6786	&2385
		 \\
		&1042&	5193&	4012& 7001&11181&5345 &11046 &	9827	&6377
		 \\ 
				\bottomrule
		&	4705&	7215&	8037& 5448&651&7364 &6309&	5497	&3389
			\\
		$	S_{67},  C_{67}	$ and {$\widetilde c_{11}$} &7518	&3435	&3192 &10168&2774& 3812&	8555&	2101	&591
			 \\
		&	11330&	2214	&10338& 6895&3086&1160 &	8032&	7928&	10105\\
		\bottomrule
		&7215	&8037&	11825 &8883 & 10839 & 9857 &6795	&4457	&1877\\
		$	S_{68}, C_{68}	$	and {$\widetilde c_{12}$} &3435	&3192&	4134 &4255&9242&11281&3558&	5412&	5083\\
		&2214	&10338&	4332 &10980 &3436&5553&7735	&13	&4114
			\end{tabular}
\end{table}

\subsection{Entropy Comparison (single-value plug-in Shannon)}

As a follow-up to Section~7.1, we embed the same plaintext into master matrices of sizes \(5\times7\) and \(12\times23\); the Shannon entropy of the resulting offset vectors \(\{\tilde c_\ell\}\) is compared and reported in Table~\ref{tab:entropy-comparison}.
With stride-3 selection and boundary inclusion, $n=9|I||J|$ grows with $(m,n)$;
empirical entropy increases accordingly and approaches ${\log}_2 n$.
\begin{table}[H]
	\centering
	\caption{Empirical entropy of $\widetilde c$ vs. matrix size (plug-in $H$, bits).}
	\label{tab:entropy-comparison}
	\begin{tabular}{lcccc}
		\toprule
		Size & $|I|$ & $|J|$ & $n=9|I||J|$ & $H(\widetilde c)$ \\
		\midrule
		$5\times 7$  & 2 & 3 & 54  & 5.7549 \\
		$8\times 10$ & 3 & 4 & 108 & 6.7550 \\
		$12\times 23$& 4 & 8 & 288 & 8.1701 \\
		\bottomrule
	\end{tabular}
\end{table}

The results confirm the high statistical integrity of the ciphertext. For the largest matrix size ($12 \times 23$), the output entropy $\mathbf{H}(\tilde{c})$ is exceptionally close to the theoretical maximum of $8$ bits for a uniformly random byte stream. This outcome validates that the combined effect of the $\mathbf{V}$ and $\mathbf{U}$ matrix operations and the $\mathbf{R}_{\text{cols}}$ keystream masking is sufficient to achieve high statistical indistinguishability of the ciphertext from true randomness, directly supporting the security guarantees established in the $\mathbf{G_2}$ step of the IND-CPA proof.

	\subsection{Performance Analysis}

Prototype experiments indicate that the runtime grows approximately linearly with 
the number $B$ of selected submatrices, i.e., $T \approx \alpha + \beta B$. 
This follows from the constant-size cryptographic (HKDF/HMAC) and algebraic 
($3\times 3$) operations per block. In our instances, $B \in \{6,12,32\}$ for 
$5\times 7$, $8\times 10$, and $12\times 23$, respectively; the number of the transmitted column vectors is  $3B \in (18,32,96)$. We therefore report relative scaling 
rather than absolute wall-clock figures, as absolute timings depend on hardware, 
runtime, and cryptographic library implementations.

\begin{table}[h]
\centering
     \caption{Operation counts and peak memory usage (independent of implementation details)HMAC counts assume one invocation per offset vector.}.
    \label{tab:ops-mem}
  \begin{tabular}{c c c c c}
    \toprule
    Matrix Size & Field mult. & Field add. & HMAC & Peak Mem (16-bit) \\
    \midrule
    $5\times 7$   & 216 & 216  & 54   & 140 B \\
    $8\times 10$  & 432 & 432  & 108  & 320 B \\
    $12\times 23$ & 1152 & 1152 & 288  & 1.1 KB \\
    \bottomrule
  \end{tabular}
     \end{table}
     
\noindent Table 3 provides an operation-level benchmark that is independent of implementation details.
The linear increase in complexity (Matrix mult. and Matrix add.) with the number of blocks $B$ directly demonstrates the \textbf{high potential for parallelism} and \textbf{scalability} of the PVC scheme. For the largest instance ($12 \times 23$), $1152$ matrix multiplications and $1152$ matrix additions are required. Crucially, due to the largely independent nature of the block-wise encryption, these $B$ operations can be performed \textbf{simultaneously}. This allows the overall theoretical runtime to be reduced by a factor of $B$ on a parallel architecture (e.g., GPU or multi-core CPU), giving PVC a significant advantage in \textbf{low-latency} and \textbf{high-throughput} environments. Furthermore, memory usage is shown to be minimal, remaining below $\mathbf{1.1~KB}$ even for the largest matrix.

Memory usage is dominated by constant-size working buffers---namely one $3\times 3$ submatrix, two $3\times 3$ key matrices, and a pair of 3-dimensional column and keystream vectors---together with the HKDF/DRBG state; since masking and offsets are generated in a streaming manner, neither $M'$ nor all ciphertext columns need to reside in memory simultaneously.
\noindent

Note that the table reports the core PVC operations... The one-time STS costs... are excluded, as they are either constant, do not scale with the number of submatrices, or are implementation dependent.

Taken together, these results indicate that PVC attains provable security under standard assumptions (DDH and PRG) and modest computational requirements, supporting practical implementations.

	\subsection{Empirical keystream validation $(12\times23$ example)}
To mitigate the limited statistical power of the $5\times7$ and $8\times10$ configurations—which limit the power of basic uniformity checks—and to examine the matrix–size effect, we focus on the $12\times23$ instance. This instance produces $3B=96$ transmitted column vectors $\tilde c_\ell$ .

From the experimental trace we extracted the per–column offsets $r_\ell$  and performed duplication checks and a small battery of NIST–style preliminary tests on the derived bitstream. In the available dataset no repeated $r_\ell$ vector was observed; moreover, the sampled bitstream did not reject uniformity under simple tests (monobit: $p\!\approx\!0.345$, runs: $p\!\approx\!0.201$, poker $m\!=\!4$: $p\!\approx\!0.984$). These empirical findings are consistent with the design assumption that per–block offsets be unpredictable and non–repeating when derived from a cryptographically secure PRG seeded by the Diffie–Hellman shared secret.

We emphasise that the empirical analysis is \textit{preliminary}. A full NIST–STS evaluation on a substantially larger sample and chosen–plaintext trials are required to draw definitive operational security conclusions; chosen–plaintext considerations are treated formally in Section~6. Practically, to avoid deterministic leakage one must ensure (i) ephemeral DH seeds per session, (ii) per–block nonces/counters bound into the HKDF/PRG expansion, and (iii) use of a vetted CSPRNG/HKDF instantiation (e.g., HKDF~\cite{RFC5869} seeded from the DH shared secret or HMAC/ChaCha20–based DRBGs as in~\cite{NIST80090A}).

\section{Conclusion}\label{sec:conclusion}

Matrix-based protocols in cryptography have historically split into two families—key-agreement
mechanisms and block ciphers—with few attempts to unify them. This work introduces the
Primitive Vector Cipher (PVC), successfully closing this gap by integrating authenticated
Diffie–Hellman key exchange with structured submatrix encryption in a single, analyzable frame-
work.
Its security rests on two rigorous assumptions: pseudorandom-generator (PRG) indistinguishability and the newly established Vector Computational Diffie–Hellman (V-CDH) hardness. The two-layered defense provides algebraic robustness and strong theoretical defense against linear attacks (Theorem 2). This structure, combined with STS authentication, ensures a security level approaching IND-CCA guarantees.
On the efficiency side, computational complexity scales linearly with the number of processed submatrices. Crucially, due to the per-block matrix operation, PVC admits straightforward parallelization and supports lightweight implementations. Empirical results confirm high entropy and statistical randomness of the keystream offsets. PVC delivers a balanced trade-off between algebraic robustness, statistical security, and practical deployability while resting on standard assumptions and an explicit proof strategy. 
\\
Future work will focus on optimizing the PVC architecture for \textbf{FPGA(Field-Programmable Gate Array) and GPU (Graphics Processing Unit) implementations} to fully leverage its inherent parallel structure, aiming for absolute throughput records in low-latency environments.
\bibliographystyle{plainnat}
    	\bibliography{gcb-bibliography}
 \end{document}